# Microrheology of viscoelastic solutions studied by magnetic rotational spectroscopy

Jean-François Berret

Matière et Systèmes Complexes, UMR 7057 CNRS
Université Denis Diderot Paris-VII, Bâtiment Condorcet, 10 rue Alice Domon et Léonie Duquet, F-75205 Paris, France, Fax: +33 1 57 27 62 11, E-mail: jean-francois.berret@univ-paris-diderot.fr

**Abstract:** Magnetic rotational spectroscopy is based on the use of magnetic micron-size wires for viscosity measurements. Submitted to a rotational magnetic field with increasing frequency, the wires undergo a hydrodynamic instability between a synchronous and an asynchronous regime. From a comparison between predictions and experiments, the static shear viscosity and elastic modulus of wormlike micellar solutions are here determined. The values agree with the determination by cone-and-plate rheometry.

## 1  Introduction

Rheology is the study of how complex materials flow and deform under stress. Traditional rheometers measure the frequency-dependent linear viscoelastic relationship between strain and stress on milliliter-scale samples. Microrheology in contrast measures these quantities using colloidal probes embedded in a fluid [1]. Fluids produced in tiny amounts or confined in small volumes, down to 1 picoliter can be examined only with this technique. The past 20 years have seen increasingly rapid advances in this research field, both theoretically and experimentally [2]. In microrheology, the objective is to translate the motion of a probe particle into the relevant fluid rheological quantities, such as the static viscosity or the high frequency elastic modulus. Following the pioneer work by Crick and Hughes some 60 years ago [3], recent studies have shown that microrheology based on the tracking of anisotropic probes, such as rods, wires and ellipsoids could bring significant advances to the field. It has been proposed that the static viscosity can be determined from the motion of a micro-actuator submitted to a rotating electric or magnetic field. These techniques are described as electric or magnetic rotational spectroscopy [4]. Here, we propose an innovative method that goes beyond the previous analysis and that is able to deal with complex fluids showing both viscosity and elasticity effects. The method is based on the tracking of magnetic wires submitted to a rotational magnetic field at increasing frequencies, and on their temporal trajectory analysis. In this work, the method is implemented on a surfactant wormlike micellar solution that behaves as an ideal Maxwell fluid. It is shown that the viscosities between 1 and $10^5$ times the water viscosity ($10^{-3}$ Pa s – $10^2$ Pa s) can be measured.

## 2   Materials and Methods

Iron oxide nanoparticles were synthesized by co-precipitation of iron(II) and iron(III) salts in aqueous media and by further oxidation of the magnetite ($Fe_3O_4$) into maghemite ($\gamma$-$Fe_2O_3$) [5,6]. The particle size and dispersity were determined from transmission electron microscopy ($D_{TEM}$ = 13.2 nm, $s_{TEM}$ = 0.23), whereas the structure was assessed by electron beam diffraction. Light scattering was used to measure the weight-average molecular weight ($M_W$ = 12×10$^6$ g mol$^{-1}$) and the hydrodynamic diameter ($D_H$ = 27 nm) of the uncoated particles [5,7]. The wires were made according to a bottom-up co-assembly process using poly(acrylic) acid coated particles together with cationic polymers [8,9]. The polymer used was poly(diallyldimethylammonium chloride) of molecular weight < 100000 g mol$^{-1}$. The wires obtained were characterized by a median length of 10 µm and a median diameter of 0.7 µm. The wires dispersion was autoclaved at 120 °C and atmospheric pressure to prevent bacterial contamination and stored at 4 °C.

**Figure 1**   a) Top and side views of the rotating field device used this work. b) Magnetic field distributions are shown along the four-coil device X and Y-directions. In the center, the magnetic field is constant over a 1×1 mm$^2$ area. c) Snapshots of a 10 µm rotating wire. The time scale between two images is 46 ms.

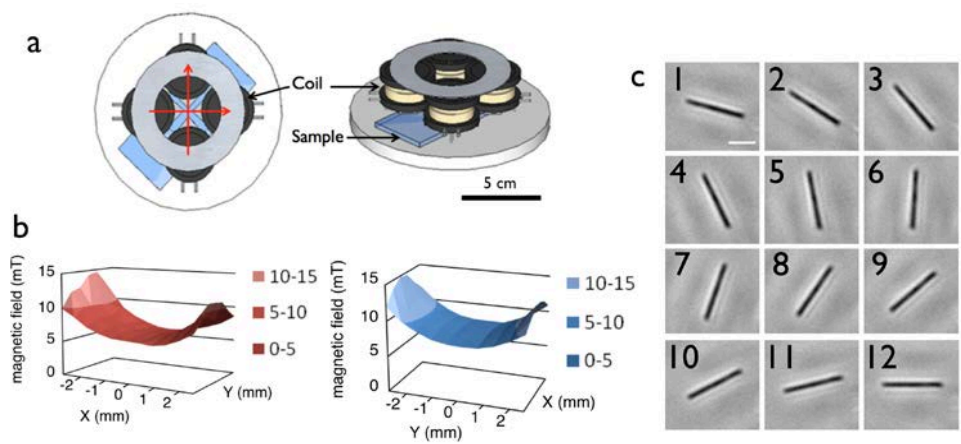

Bright field microscopy was used to monitor the wire actuation as a function of time. Stacks of images were acquired on an IX73 inverted microscope (Olympus) equipped with a 100× objective. 65 µl of a fluid to be studied were deposited on a glass plate and sealed into to a Gene Frame® (Abgene/Advanced Biotech) dual adhesive system. To prevent wire rotation coupling and interaction, experiments were performed in infinite dilution, corresponding to wire concentrations of 1 pM. The glass plate was introduced into a homemade device generating a rotational magnetic field, thanks to two coil pairs working with a 90°-phase shift (Fig. 1a). An electronic set-up allowed measurements in the frequency range 10$^{-3}$ - 10 rad s$^{-1}$ and at magnetic fields B = 0 – 20 mTesla [10]. Fig. 1b displays the magnetic field distributions between the electromagnetic coil poles in the X- and Y-directions. The image acquisition system consisted of an EXi Blue CCD camera (QImaging) working with Metaview (Universal Imaging). Images of wires were digitized and treated by the ImageJ software and plugins. Fig. 1c shows snapshots of a rotating 10 µm wire at fixed time interval during a 180° spin.

## 3   Results and Discussion

*Newtonian liquid* - For a viscous liquid of viscosity $\eta$, a wire submitted to a rotating field experiences a restoring torque that slows down its rotation. As a result, the wire motion undergoes a transition between a synchronous and an asynchronous rotation. In the first regime, the wire rotates at the same frequency as the field, whereas in the second regime it is animated of periodic back-and-forth motions. Because wires are superparamagnetic, the oscillation frequency above the instability is twice that of the excitation. The critical frequency $\omega_C$ between these two regimes reads [10-12]:

$$\omega_C = \frac{3}{8}\frac{\mu_0 \Delta\chi}{\eta_0} g\left(\frac{L}{D}\right)\frac{D^2}{L^2} H^2 \qquad (1)$$

where $\mu_0$ is the permeability in vacuum, $L$ and $D$ the length and diameter of the wire, $H$ the magnetic excitation amplitude and $g(L/D) = Ln(L/D) - 0.662 + 0.917 D/L - 0.050(D/L)^2$ is a dimensionless function of the anisotropy ratio. In Eq. 1, $\Delta\chi = \chi^2/(2 + \chi)$ where $\chi$ denotes the magnetic susceptibility. In the synchronous and asynchronous regimes, the wire motion is analyzed with respect to their rotation angle $\theta(t)$. Examples of $\theta(t)$-traces obtained are illustrated in Fig. 2. For the analysis, the $\theta(t)$-traces are translated into a set of two parameters: the average rotation velocity $\Omega(\omega)$ and the oscillation amplitude $\theta_B(\omega)$ in the asynchronous regime [10]. The average angular velocity $\Omega(\omega)$ in the two regimes expresses as:

$$\begin{aligned}\omega \leq \omega_C \quad & \Omega(\omega) = \omega \\ \omega \geq \omega_C \quad & \Omega(\omega) = \omega - \sqrt{\omega^2 - \omega_C^2}\end{aligned} \qquad (2)$$

With increasing frequency, the average velocity increases linearly, passes through a cusp-like maximum at the critical frequency and decreases. The transition between the synchronous and asynchronous regimes was used to calibrate the wire-rheometer and determine the susceptibility parameter $\Delta\chi$ in Eqs. 1. For the calibration, experiments were performed on a 85 wt. % water-glycerol mixture of static viscosity $\eta$ = 0.062 Pa s$^{-1}$. In a purely viscous fluid, the critical frequency is found to decrease as $g\left(\frac{L}{D}\right)L^{-2}$ in agreement with Eq. 1. For wires made from 13.2 nm particles and PADADMAC, we found $\Delta\chi$ = 2.1 ± 0.4, and a magnetic susceptibility $\chi$ = 3.4 ± 0.4.

*Maxwell fluid* – In continuum mechanics, a Maxwell fluid is described by a spring and dashpot in series. An actuated wire immersed in such a medium experiences a viscous and an elastic torque that both oppose the applied magnetic torque. The differential equation describing the wire motion has been derived and solved, leading to the following predictions [10]. With increasing ω, the wire undergoes the same type of transition as the one described previously and the critical frequency $\omega_C$ expresses as in Eq. 1. The static viscosity $\eta$ in Eq. 1 is however replaced by the product $G\tau$, where $G$ and $\tau$ denote the shear elastic modulus and the fluid relaxation time. The set of equations in Eq. 2 is also identical to that of a Newtonian fluid. From the oscillation amplitude $\theta_B(\omega)$ in the asynchronous regime, it is possible to determine the shear elastic modulus $G$ using [10]:

$$\lim_{\omega \to \infty} \theta_B(\omega) = \frac{3}{4} \frac{\mu_0 \Delta \chi}{G} g\left(\frac{L}{D}\right) \frac{D^2}{L^2} H^2 \tag{3}$$

***Comparison between cone-and-plate and wire-based rheology*** – To assess the validity of the wire-based microrheology approach, magnetic wires were tested on a surfactant solution made of cetylpyridinium chloride and sodium salicylate, abbreviated as CPCl/NaSal. CPCl and NaSal are known to self-assemble spontaneously into micrometer long wormlike micelles, which then build a semi-dilute entangled network above a certain concentration [13,14]. This network confers to the solution a Maxwell–type viscoelastic behavior. At the concentration of 2 wt.%, the network mesh size is of the order of 30 nm, *i.e.* much smaller than the wire diameter. Surfactant wormlike micelles are sometimes called equilibrium polymers.

In Fig. 2, a 15 mT rotating magnetic field was applied to a 8 µm nanowire immersed in the CPCl/NaSal solution at increasing frequencies: a) $\omega = 0.85$ rad s$^{-1}$; b) $\omega = 1.2$ rad s$^{-1}$ and c) $\omega = 7$ rad s$^{-1}$. The wire motion was monitored by optical microscopy, and the time dependence of their orientation was derived. At low frequency, the wire rotates with the field and $\theta(t) = \omega t$. Above the critical frequency (here $\omega_C = 0.90$ rad s$^{-1}$), the wire is animated of back-and-forth motion characteristic of the asynchronous regime. In this range, $\theta(t)$ displays oscillations. The green straight lines in the figure represent the average angular velocity $\Omega(\omega)$ introduced in Eq. 2.

**Figure 2**   Rotation angle $\theta(t)$ of a 8.1 µm wire as a function of the time at various frequencies: a) $\omega = 0.8$ rad s$^{-1}$; b) $\omega = 1.2$ rad s$^{-1}$ and c) $\omega = 7$ rad s$^{-1}$. The behaviour in a) corresponds to the synchronous regime. The green straight lines in b) and c) represent the average angular velocity $\Omega = d\theta/dt$ in the asynchronous regime.

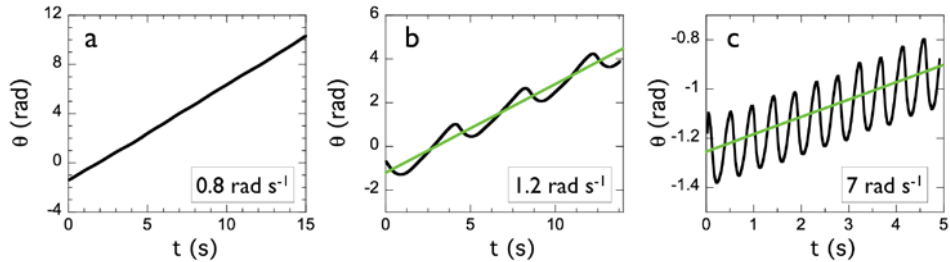

Fig. 3a displays the frequency dependence of the elastic and loss moduli $G'(\omega)$ and $G''(\omega)$ obtained on the 2 wt.% CPCl/NaSal solution. Data were obtained on a CSL 100 rheometer from TA Instruments with a cone-and-plate device (inset). At $T = 27$ °C, this wormlike micellar fluid was characterized by the static viscosity $\eta = 1.0$ Pa s, a relaxation time $\tau = 0.14$ s and an elastic modulus $G = 7.1$ Pa, in agreement with earlier literature [13,14]. In Fig. 3a, the $G'(\omega)$ and $G''(\omega)$ data were normalized with respect to the elastic modulus and the frequency with respect to the relaxation time. The continuous lines are the predictions for a Maxwell fluid: $G'(X)/G = X^2/(1 + X^2)$ and $G''(\omega)/G = X/(1 + X^2)$, where $X = \omega\tau$. The agreement between the data and predictions is excellent.

**Figure 3** a) Cone-and-plate geometry used in shear rheometry. $G'(\omega)/G$ and $G''(\omega)/G$ versus frequency for a wormlike micellar fluid. The continuous lines are Maxwell predictions. Inset: cone-and-plate device used in rheometry. b) Average angular velocity $\Omega(\omega)$ as a function of the frequency. The solid line corresponds to the best fit using Eq. 1. Inset in b): image of the 8.1 μm wire by phase contrast microscopy (60×, scale bar 5 μm).

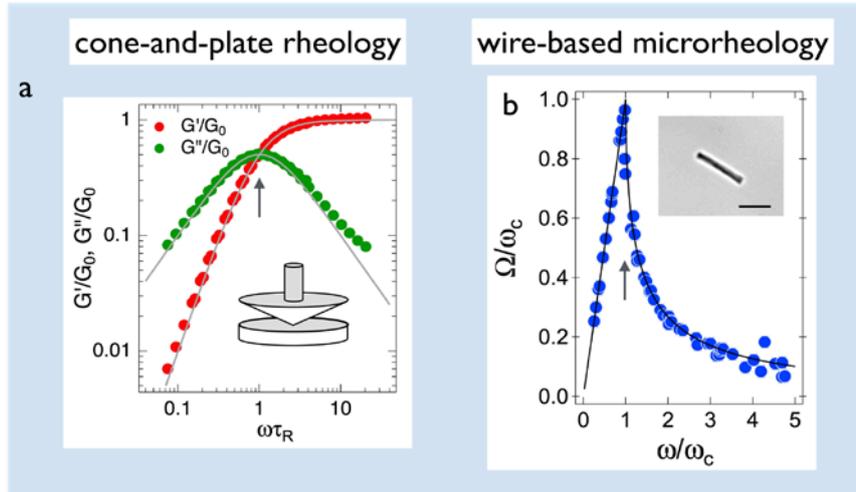

**Table 1** Viscoelastic parameters of a CPCl/NaSal wormlike micellar solution at 2 wt. % using magnetic rotational spectroscopy.

| Wire | #1 | #2 | #3 | #4 | #5 | #6 | #7 | #8 | #9 |
|---|---|---|---|---|---|---|---|---|---|
| $L$ (μm) | 8.2 | 14.8 | 8.1 | 8.1 | 12.5 | 5.0 | 3.3 | 6.3 | 3.0 |
| $B$ (mT) | 10.4 | 10.4 | 10.4 | 15.3 | 7.3 | 14.8 | 14.8 | 10.4 | 14.8 |
| $\omega_C$ (rad s$^{-1}$) | 1.44 | 0.55 | 0.38 | 0.88 | 0.41 | 1.6 | 4.0 | 0.80 | 2.1 |
| $\eta$ (Pa s) | 1.11 | 1.21 | 1.45 | 1.32 | 0.97 | 1.49 | 1.08 | 1.37 | 1.72 |
| $\tau$ (s) | 0.15 | 0.13 | 0.16 | 0.14 | 0.12 | 0.14 | 0.12 | 0.17 | 0.15 |
| $G$ (Pa) | 7.6 | 9.5 | 9.2 | 9.7 | 7.9 | 10.3 | 9.1 | 7.9 | 11.2 |

Fig. 3b displays the average angular velocity $\Omega(\omega)$ *versus* ω obtained for a 8.1 μm wire. With increasing frequency, $\Omega(\omega)$ increases, passes through a maximum at the critical frequency and decreases. The $\Omega(\omega)$-data were adjusted using Eq. 2 and a value of the static viscosity of $\eta = 1.3 \pm 0.3$ Pa s, in good agreement with the cone-and-plate rotational rheometry value. From the oscillation amplitudes in the regime $\omega\tau \gg 1$ (Eq. 3), a modulus of $9.4 \pm 2$ Pa was derived. Experiments performed with wires of different lengths and in various magnetic field conditions confirmed the good agreement with cone-and-plate rheometry, and demonstrates the ability of the technique to measure the linear viscoelasticity of a Maxwell fluid (Table 1).

## 4 – Conclusion

We have shown here that micron-size wires with magnetic properties can be used as probes for active µ-rheology. The magnetic rotational spectroscopy studies the transition between a synchronous and an asynchronous regime of rotation observed with increasing frequency. The instability resembles that observed on laboratory benches when a viscous solution is actuated with a magnetic bar. Tested here on a Maxwell fluid, viscoelastic parameters $\eta$, $G$ and $\tau$ of the fluid can be derived. The technique can be extended in principle to any kind of complex fluids.

## Acknowledgements

I am grateful to L. Chevry, A. Cebers, M.-A. Fardin and A. Hallou for fruitful discussions. The master students who participated to this research, C. Leverrier, A. Conte-Daban, C. Lixi, L. Carvhalo and R. Chan are also acknowledged.